\documentclass[amsmath,twocolumn]{aastex702}
\usepackage{CJK}
\usepackage{booktabs,float,bm}

\newcommand{\LCDM}{$\Lambda$CDM}
\newcommand{\wdm}{w_{\rm DM}}
\newcommand{\wde}{w_{\rm DE}}
\newcommand{\CPL}{$w_0w_a$}
\newcommand{\DICV}{\mathrm{DIC}_{V}}

\begin{document}
\begin{CJK*}{UTF8}{gbsn}

\title{Is Dark Matter Really Matter?}

\author[orcid=0000-0003-0957-3633,sname='Jiang']{Jun-Qian Jiang (江俊钱)}
\affiliation{Korea Astronomy and Space Science Institute, Daejeon 34055, Korea}
\email[show]{jiang@kasi.re.kr}  

\author[orcid=0000-0001-6815-0337,sname='Shafieloo']{Arman Shafieloo} 
\affiliation{Korea Astronomy and Space Science Institute, Daejeon 34055, Korea}
\affiliation{University of Science and Technology, Daejeon 34113, Korea}
\email[show]{shafieloo@kasi.re.kr}

\begin{abstract}

In the standard model of cosmology, it is assumed that dark matter is pressureless with equation of state $w=0$ and dark energy has $w=-1$.
We test these assumptions jointly using DESI DR2 distance measurements,
including the recent Lyman-$\alpha$ Alcock--Paczynski (AP)
information, DES supernovae, and two complementary CMB treatment.
When constant $w_{\rm DM}$ and $w_{\rm DE}$ are varied together, we find
$w_{\rm DM}=0.000968^{+0.000501}_{-0.000496}$ and
$w_{\rm DE}=-0.9380^{+0.0259}_{-0.0262}$ (68\%).
With an alternative CMB treatment that marginalizes over the lensing spectrum, the
corresponding constraints are
$w_{\rm DM}=0.000870^{+0.000408}_{-0.000410}$ and
$w_{\rm DE}=-0.9353^{+0.0258}_{-0.0254}$.
Both standard $\Lambda$CDM values are disfavored at approximately $2\sigma$ in the joint
extension.
Neither parameter departs significantly from its standard value when only
that parameter is varied.
This behavior arises because late-time distances favor $w_{\rm DE}>-1$, while
maintaining the
early-Universe physical matter density
requires a compensating positive $w_{\rm DM}$,
which changes the mapping to the matter density today.
Allowing dynamical dark energy further highlights the complexity of the situation: phantom crossing for dark energy makes $w_{\rm DM}=0$ consistent with the data, whereas a positive $w_{\rm DM}$ preference persists when crossing is forbidden.
Interestingly, the Pad'e-$w$ parameterization that provides a flexible description of a class of quintessence models (with no phantom crossing), along with $w_{\rm DM}$ free, is even mildly favored over the phantom-crossing $w_0w_a$ model according to both the best-fit $\chi^2$ and the DIC under both CMB treatments. One can conclude that the apparent preference for phantom crossing may instead reflect deviations in the dark-matter sector rather than dark-energy dynamics alone.
With future high-redshift distance measurements, particularly improved Lyman-$\alpha$ observations, together with growth and lensing probes, these possibilities may be further clarified.

\end{abstract}


\section{Introduction}
\label{sec:introduction}

The \LCDM\ model provides a remarkably successful description of a broad
range of cosmological observations, from the cosmic microwave background
to large-scale structure and Type Ia supernovae.  In its standard
formulation, however, the dark sector is specified by two exact assumptions:
dark matter is a pressureless fluid with $\wdm=0$, while cosmic acceleration
is attributed to a cosmological constant with $\wde=-1$.  Neither assumption
is guaranteed \emph{a priori}; both are hypotheses to be tested against
increasingly precise data.

Recent observational developments have further sharpened interest in going beyond the standard dark energy sector.
In particular, DESI BAO measurements, especially when combined with CMB and Type Ia supernova data, have stimulated extensive discussion of phenomenological models of evolving dark energy beyond $w_{\rm DE} = -1$ \citep{DESI:2024mwx,DESI:2025zgx,DESI:2024aqx,DESI:2024kob,DESI:2025fii,Li:2026hwq,Li:2026asg}.
Such indications are especially intriguing because the preferred evolution may involve crossing the phantom divide, a behavior that is nontrivial to realize in theoretically consistent single-field descriptions and often calls for additional degrees of freedom or more general constructions (see e.g.~\citep{Vikman:2004dc,Caldwell:2005ai,Fang:2008sn}).
However, strong degeneracies can exist among dark-sector properties~\citep{Shafieloo:2011zv,Kunz:2007rk}, and an apparent time-dependent deviation in dark energy may instead be a projection effect induced by departures of dark matter properties from their $\Lambda$CDM values.
Indeed, even the separation between dark matter and dark energy can itself be questioned~\citep{Kou:2025yfr}.
In this context, it is natural to ask whether the dark matter sector should still be kept fixed to its exact cold, pressureless limit while only the dark energy sector is generalized, or whether both standard assumptions in the dark sector should be examined on an equal footing.

Early constant-$\wdm$ constraints from CMB, BAO,
supernova, and large-scale-structure data were consistent with pressureless
dark matter~\citep{Muller:2004yb,Xu:2013mqe,Thomas:2016iav}, and analyses varying constant $\wdm$ and $\wde$
together reached the same conclusion~\citep{Kumar:2012gr}.  Generalized
dark matter~\citep{Hu:1998kj}, redshift-binned reconstructions
~\citep{Kopp:2018zxp,Ilic:2020onu,Zhumabek:2023ejd}, and later variable-$\wdm$ and nonzero-sound-speed
analyses~\citep{Yadav:2023qfj} broadened these tests without compelling
evidence against the cold limit.  Noncold or barotropic dark matter has
also been combined with nonstandard dark-energy histories
~\citep{Yao:2023ybs,Liu:2025mob}.
Recently, a late transition away from dust has been shown to
reproduce expansion histories statistically comparable to \CPL, with
interpretations ranging from backreaction to exotic dark matter
~\citep{Giani:2025hhs,Braglia:2025gdo}.  Analyses that instead keep a
constant $\wdm$ find that its preference can strengthen for constant
$\wde$ and weaken once unrestricted \CPL\ evolution is admitted
~\citep{Li:2025dwz}. In phenomenologically emergent dark energy, the
preferred sign and model evidence are different~\citep{Li:2025eqh}.

Motivated by the latest DESI BAO and AP measurements~\citep{DESI:2025zgx,DESI:2026lnd}, we ask what happens
when the equations of state of dark matter and dark energy are varied
simultaneously.  To reduce sensitivity to uncertain nonlinear matter-power
modeling, we adopt two complementary CMB likelihood constructions.  Our
analysis covers the constant-$w$ $ww$DM model and its two one-parameter
restrictions, and extends to dynamical dark energy through unrestricted and
nonphantom \CPL\ forms and a nonphantom Pad\'e parameterization well suited
to describing quintessence evolution.  In the
constant-$w$ model, the standard values of both equations of state are
disfavored at approximately $2\sigma$, while neither shows a comparable
departure when varied alone.  With dynamical dark energy, the preference
for positive $\wdm$ depends strongly on whether phantom crossing is allowed.
In particular, the nonphantom Pad\'e-$w$+$\wdm$ model is favored over
unrestricted \CPL\, in the model comparison for both CMB treatments.

The paper is organized as follows.
Section~\ref{sec:models} defines the models.
Section~\ref{sec:data} describes the data and inference.
The constant and dynamical dark energy results are presented in
Secs.~\ref{sec:constant} and~\ref{sec:dynamical}.
Section~\ref{sec:comparison} compares the model fits, and
Sec.~\ref{sec:conclusions} concludes.
Appendix~\ref{app:priors-posteriors} gives the priors and full posterior
constraints, Appendix~\ref{app:alternative-sn} tests alternative supernova
samples, and Appendix~\ref{app:nonlinear} gives the nonlinear diagnostic.
Appendix~\ref{app:lya-ap-validation} assesses the applicability
of the Lyman-$\alpha$ AP compression to our analysis.

\section{Dark-sector models}
\label{sec:models}

\subsection{Constant equations of state}

We model dark matter and dark energy as noninteracting fluids whose background density $\bar\rho_i$ obeying
\begin{equation}
 \bar\rho_i(a)=\bar\rho_{i}(a=1)a^{-3(1+w_i)} .
\label{eq:densityconstant}
\end{equation}
Our constant model, denoted $ww$DM, varies both their equations of state $\wdm$ and
$\wde$.  The restrictions $\wde=-1$ and $\wdm=0$ are denoted
$\Lambda w$DM and $w$CDM, respectively.  Setting both to their standard
values gives \LCDM.

For a general fluid, the synchronous-gauge density and velocity
perturbations satisfy~\citep{Ma:1995ey}
\begin{align}
\dot\delta={}&-(1+w)\left(\theta+\frac{\dot h}{2}\right)
-3{\cal H}(c_s^2-w)\delta \nonumber\\
&-9{\cal H}^2(1+w)(c_s^2-c_a^2)\frac{\theta}{k^2},\\
\dot\theta={}&-{\cal H}(1-3c_s^2)\theta+
\frac{c_s^2}{1+w}k^2\delta ,
\end{align}
where $c_a^2=\dot{\bar P}/\dot{\bar\rho}$ and anisotropic stress is set to
zero.  We fix the dark-matter rest-frame sound speed to
$c_{s,\rm DM}^2=0$.  Dark-energy perturbations are treated with
$c_{s,\rm DE}^2=1$.  The models are implemented in
\texttt{CLASS}~\citep{Blas:2011rf}.

\subsection{Dynamical dark energy and phantom crossing}

Our first dynamical form for $w_{\rm DE}$ is \CPL (Chevallier--Polarski--Linder, CPL)~\citep{Chevallier:2000qy,Linder:2002et},
\begin{equation}
 w_{\rm DE}(a)=w_0+w_a(1-a).
\label{eq:cpl}
\end{equation}
We analyze both unrestricted \CPL\ and \CPL+$\wdm$.
We then consider the case with a nonphantom prior:
\begin{equation}
 w_0>-1,\qquad w_0+w_a>-1.
\label{eq:nonphantom}
\end{equation}
The restriction is motivated by the physical difficulties associated with
phantom crossing and the relative simplicity of realizing quintessence dark
energy.

We also use the two-parameter Pad\'e-$w$ form, which provides a flexible
description of quintessence evolution,
\begin{equation}
 w_{\rm DE}(a)=-1+
 \frac{2\epsilon_0 a^3}{(3-\eta_0)a^3+\eta_0},
\label{eq:pade}
\end{equation}
with $0<\epsilon_0<3$ and $0<\eta_0<100$.
These priors keep the sampled histories on the nonphantom side.
It typically remains well behaved over a wider redshift range.
We consider Pad\'e-$w$ with either fixed or free $\wdm$.

\section{Data and inference}
\label{sec:data}

We combine three classes of observations.  For late-time distances we use
the DESI DR2 BAO measurements from the BGS, LRG, ELG, and quasar tracers
~\citep{DESI:2025zgx}.
At $z_{\rm eff}=2.33$ we use the recent joint BAO+AP compression~\citep{DESI:2026lnd}.
The compression retains the
geometric information in $D_M/r_d$ and $D_H/r_d$ while marginalizing the
smooth isotropic broadband scale and small-scale contamination.
Since the full-shape AP measurement is extracted using a fiducial \LCDM\ power-spectrum template, we discuss in Appendix~\ref{app:lya-ap-validation} why it remains applicable to our models.
We use the DES 5-year Dovekie Type Ia supernova data~\citep{DES:2025sig}.
We also tested other supernova data in Appendix~\ref{app:alternative-sn}.

These dark-sector models do not yet have a simulation-calibrated nonlinear
matter-power prescription.  Applying a standard \LCDM-calibrated correction
would therefore propagate an uncontrolled late-time prediction into CMB
lensing.
We address this limitation with two
complementary CMB constructions: one removes the more nonlinear small-scale
anisotropy information, and the other retains high signal-to-noise while
marginalizing over the lensing-potential shape.

The first, denoted \emph{CMB-$1000$}, combines Planck low-$\ell$ TT (\texttt{Commander}) and EE (\texttt{SimAll})
with NPIPE CamSpec-lite TT, TE, and EE~\citep{Planck:2019nip,Rosenberg:2022sdy,Jense:2025wyg}, applying a conservative cut
$\ell_{\max}=1000$.
We use the lite likelihood because foreground and instrumental nuisance
parameters have already been marginalized into the spectrum covariance.
This is important when imposing scale cuts~\citep{Jense:2025wyg}.

The second, denoted \emph{CMB lensing-marginalized},
uses CamSpec-lite to $\ell_{\max}=1500$, $1000$, and $600$
for TT, TE, and EE, respectively.  ACT DR6 supplies the corresponding
spectra above those boundaries up to $\ell=6500$, and the SPT-3G D1 lite
likelihood is included over its full range
~\citep{AtacamaCosmologyTelescope:2025blo,SPT-3G:2025bzu}.  In the overlapping
sky and multipole ranges between Planck and ACT, these scale cuts are optimized for
signal-to-noise.
Motivated by~\cite{Lemos:2023xhs}, the theoretical
$D_L^{\phi\phi}=[L(L+1)]^2C_L^{\phi\phi}/(2\pi)$, where the nonlinear information entered, is replaced by a
log-cubic spline with six amplitudes at
\begin{equation}
 L=(7,44,125,600,1600,3100).
\end{equation}
The sampled node variables are $\ln D_L^{\phi\phi}$, each with a flat prior
$-22<\ln D_L^{\phi\phi}<-14$, and the interpolation is performed in $\ln L$.
\texttt{CLASS} supplies the unlensed primary spectra, which are relensed
with this sampled spline.  The same empirical $C_L^{\phi\phi}$ enters the
Planck+ACT+SPT four-point reconstruction likelihood through $L=4000$
~\citep{ACT:2025qjh,Carron:2022eyg,ACT:2023ubw,ACT:2023dou,ACT:2023kun,SPT-3G:2024atg}.  The six amplitudes are sampled jointly with the
cosmological, calibration, and foreground parameters and are then
marginalized.
By marginalizing over the CMB lensing spectrum, we eliminate the dependence
on nonlinear information.

We sample the posteriors with \texttt{Cobaya}~\citep{Torrado:2020dgo} and
require Gelman--Rubin $R-1<0.01$~\citep{Gelman:1992zz}.
We use BOBYQA optimizer to locate the best fit~\citep{Cartis:2018xum,Cartis:2018jxl}.

\section{Constant dark sector}
\label{sec:constant}

\begin{figure*}[htb]
\centering
\begin{minipage}[t]{0.495\textwidth}
\centering
\small CMB-$1000$\par
\includegraphics[width=\linewidth]{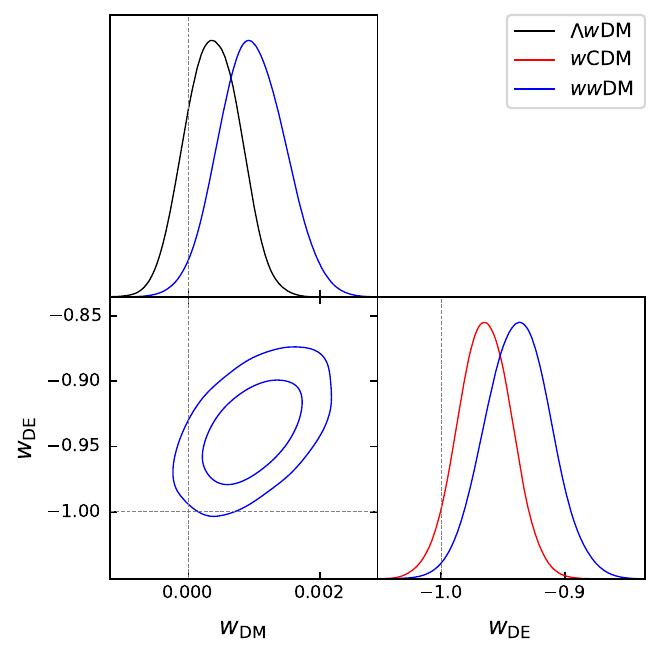}
\end{minipage}\hfill
\begin{minipage}[t]{0.495\textwidth}
\centering
\small CMB lensing-marginalized\par
\includegraphics[width=\linewidth]{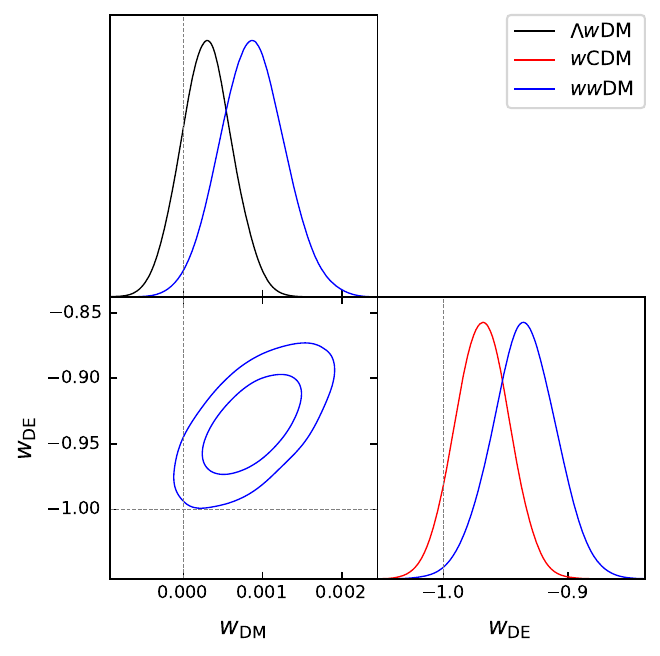}
\end{minipage}
\caption{Marginalized constraints for the constant dark-sector models.
The left triangle uses CMB-$1000$ and the right triangle uses the
CMB lensing-marginalized likelihood,
always combined with DESI DR2 and DES supernovae.
The diagonal panels show one-dimensional marginalized posteriors, while
the lower-left panels show the joint $\wdm$--$\wde$ constraints.
Contours enclose 68\% and 95\% probability, and dashed lines mark the
\LCDM\ values.  Releasing only $\wdm$ or only $\wde$ produces no significant
departure, whereas releasing both selects the correlated quadrant
$\wdm>0$, $\wde>-1$.}
\label{fig:constant}
\end{figure*}

Table~\ref{tab:constant} and Fig.~\ref{fig:constant} show the constant dark sector
results.  For CMB-$1000$, the joint model gives (68\% credible level, here and below)
\begin{align}
 \wdm&=0.000968^{+0.000501}_{-0.000496},\nonumber\\
 \wde&=-0.9380^{+0.0259}_{-0.0262},
\end{align}
Both standard values lie just beyond their 95\% marginal limits.\footnote{The corresponding 95\% constraints are
$\wdm=0.000968^{+0.000983}_{-0.000953}$ and
$\wde=-0.9380^{+0.0521}_{-0.0512}$.}
The
CMB lensing-marginalized analysis yields comparable means and slightly
smaller uncertainties.
For both CMB treatments, the \LCDM\ point also lies outside the 95\%
joint $\wdm$--$\wde$ contour.
In contrast, the $\Lambda w$DM mean is below $1\sigma$ from zero, while
the $w$CDM mean is only $1.4$--$1.6\sigma$ from $-1$.
Thus the approximately $2\sigma$ shifts are a property of the
\emph{joint} extension.
The agreement between the two CMB treatments indicates that the preference is mainly driven by the primary-CMB constraints on the early-Universe matter density and acoustic scale. The agreement between the two likelihood constructions also provides an additional robustness check that the inferred preference is not sensitive to the treatment of nonlinear lensing.

\begin{table*}[htb]
\centering
\begin{tabular}{llcc}
\toprule
CMB & model & $\wdm$ & $\wde$\\
\midrule
CMB-$1000$ & $\Lambda w$DM &
$0.000370^{+0.000417}_{-0.000424}\;(0.000358)$ &
$-1\ \mathrm{(fixed)}$\\
& $w$CDM &
$0\ \mathrm{(fixed)}$ &
$-0.9649^{+0.0215}_{-0.0214}\;(-0.9645)$\\
& $ww$DM &
$0.000968^{+0.000501}_{-0.000496}\;(0.000913)$ &
$-0.9380^{+0.0259}_{-0.0262}\;(-0.9481)$\\
\midrule
CMB lensing-marginalized & $\Lambda w$DM &
$0.000294^{+0.000324}_{-0.000323}\;(0.000425)$ &
$-1\ \mathrm{(fixed)}$\\
& $w$CDM &
$0\ \mathrm{(fixed)}$ &
$-0.9687^{+0.0216}_{-0.0216}\;(-0.9719)$\\
& $ww$DM &
$0.000870^{+0.000408}_{-0.000410}\;(0.000707)$ &
$-0.9353^{+0.0258}_{-0.0254}\;(-0.9435)$\\
\bottomrule
\end{tabular}
\caption{Constant dark-sector constraints.  Each sampled entry is the
posterior mean with 68\% equal-tail errors, followed in parentheses by the
best fit.}
\label{tab:constant}
\end{table*}

The shift along this degeneracy should be understood as jointly
driven by the BAO/AP+SN data and early-Universe information from the CMB.
\footnote{We verified that, when the CMB data are excluded, the constraint
on $\wdm$ becomes much weaker and no comparably significant preference
emerges.}
A value of $\wde$ above $-1$ provides the energy density decay desired by the late-time BAO and supernova geometry measurements.
Although a nonzero $\wdm$ can also modify the late-time expansion history, the value is too small for low-redshift tracers.

\begin{figure*}[htb]
\centering
\includegraphics[width=\textwidth]{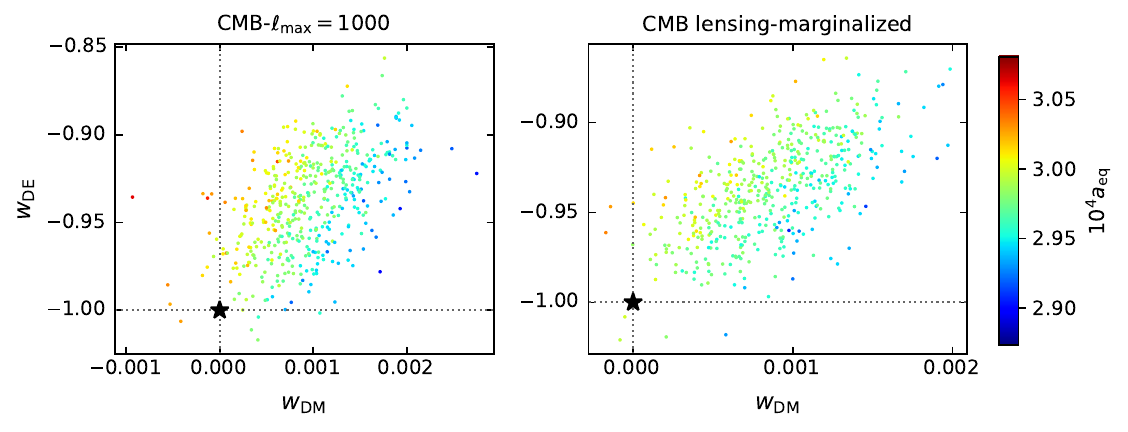}
\caption{The equality epoch in the constant $ww$DM model.
Posterior samples in the $(\wdm,\wde)$ plane are colored by
$10^4a_{\rm eq}$ for CMB-$1000$ (left) and the CMB
lensing-marginalized likelihood (right).  The star marks \LCDM.
The correlated displacement in the two equations of state occupies a
narrow equality range, equivalently a narrow early-Universe $\Omega_mh^2$.
Late-time geometry selects the $\wde$ displacement, and $\wdm$ changes the
conversion between the present and CMB-calibrated early dark-matter
densities.}
\label{fig:equality}
\end{figure*}

At the high-redshift end,
the CMB temperature and polarization spectra constrain the matter-to-radiation
ratio around last scattering through radiation driving and the early
integrated Sachs--Wolfe contribution~\citep{Hu:1995en,Hu:2001bc,Planck:2018vyg,Weiner:2026sfm}.
This information can be equivalently shown as the equality epoch $a_{\rm eq}$ in \autoref{fig:equality}.

The high-redshift background evolution can be characterized using the
matter-era distance interval (MEDI)~\citep{Weiner:2026sfm},
\begin{equation}
 {\cal D}_{\rm ME}\equiv
 \frac{D_M(z_d)-D_M(z_m)}{r_d},\qquad z_m=2.33 ,
\label{eq:medi}
\end{equation}
where $z_d$ is photon--baryon decoupling redshift.  The CMB acoustic angle measures
$D_M(z_d)/r_d$, while BAO/AP measures $D_M(z_m)/r_d$.
Their difference removes
the distance accumulated below the Lyman-$\alpha$ redshift $z_m$ and is consequently dominated by the
expansion during matter domination.
Ref.~\citep{Weiner:2026sfm,DESI:2026lnd} found the
high-redshift-agnostic acoustic-scale measurement
${\cal D}_{\rm ME}$ is higher than
\LCDM\ prediction.
In particular, the recent high-redshift Lyman-$\alpha$ measurement
included here substantially tightens its constraint.

Positive $\wdm$ changes the extrapolation entering that comparison.
At fixed density at decoupling,
\begin{equation}
 \frac{\rho_{\rm DM}(a_m)}
 {\rho_{\rm DM}^{(w=0)}(a_m)}
 =\left(\frac{a_m}{a_d}\right)^{-3\wdm}<1 .
\label{eq:medidilution}
\end{equation}
At the $ww$DM best fits, Eq.~(\ref{eq:medidilution}) implies about $1.6\%$
and $1.2\%$ less dark matter at $z_m$ than a pressureless component
extrapolated from the same decoupling density, for CMB-$1000$ and CMB
lensing-marginalized, respectively.  The reduced matter abundance lowers
the expansion rate over the interval and increases ${\cal D}_{\rm ME}$, in
the direction required to relieve the distance excess.
From the \LCDM\ to the $ww$DM best fit,
\begin{align}
 {\cal D}_{\rm ME}&: 55.377\longrightarrow55.475
 \quad\text{(CMB-$1000$)},\nonumber\\
 {\cal D}_{\rm ME}&: 55.375\longrightarrow55.489
 \quad\text{(lensing-marginalized CMB)} ,
\label{eq:medinumbers}
\end{align}
increases of $0.098$ ($0.177\%$) and $0.114$ ($0.205\%$), respectively.
This shift therefore reduces the tension with
the high-redshift-agnostic measurement (see e.g. Fig.8 in \cite{DESI:2026lnd}).

\section{Dynamical dark energy}
\label{sec:dynamical}

The correlation between $\wdm$ and $\wde$ found in the constant-$w$
analysis motivates extending the test to dynamical dark energy.  In
particular, if departures in the two dark sectors can compensate each other
in the distance--redshift relation, the apparent preference for a dark-energy
history that crosses the phantom divide may depend on the assumption that
dark matter is exactly pressureless.  We therefore ask whether allowing
$\wdm$ to vary changes the interpretation of phantom crossing, and compare
models in which such crossing is allowed or forbidden.

\begin{figure*}[htp]
\centering
\includegraphics[width=\textwidth]{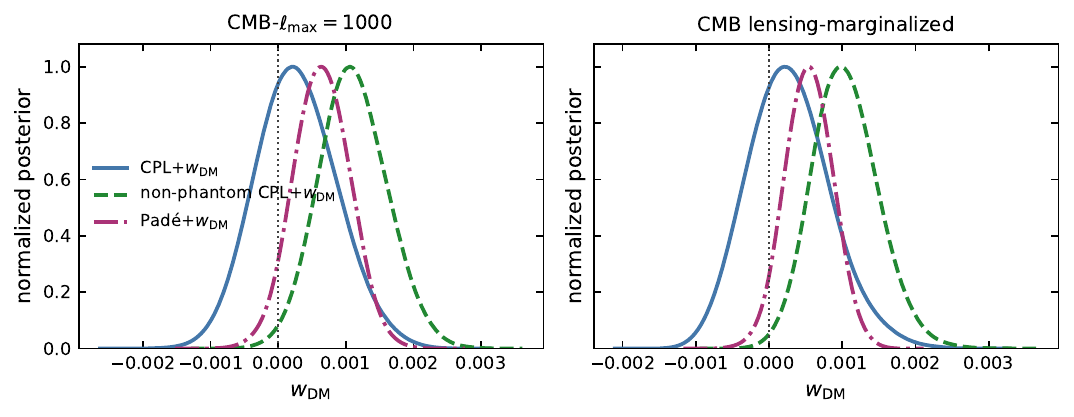}
\caption{Dynamical dark-energy results for the two CMB treatments.
The panels compare marginalized $\wdm$ posteriors for CMB-$1000$ (left)
and the CMB lensing-marginalized likelihood (right).
Unrestricted \CPL+$\wdm$ allows phantom crossing and is consistent with
$\wdm=0$.  The positive preference returns when crossing is forbidden,
both for nonphantom \CPL+$\wdm$ and for Pad\'e-$w$+$\wdm$.}
\label{fig:dynamical}
\end{figure*}

The posterior constraints and best fits for the cases with dynamical dark energy are given in
Tables~\ref{tab:dynamicwdm}.
Allowing unrestricted \CPL\ changes the inference qualitatively.
For CMB-$1000$ we find
$\wdm=0.000272^{+0.000608}_{-0.000601}$; the CMB
lensing-marginalized likelihood gives
$\wdm=0.000276^{+0.000557}_{-0.000572}$.
The freedom to cross $w=-1$ can reproduce the high-redshift distance shape.

\begin{figure*}[htb]
\centering
\includegraphics[width=\textwidth]{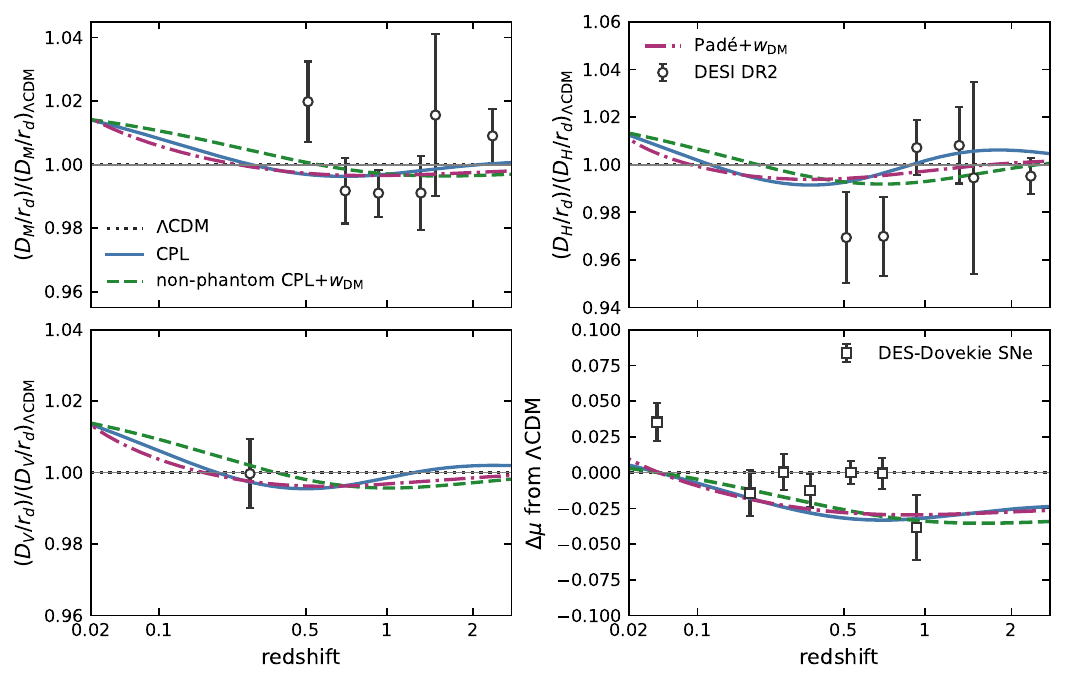}
\caption{Expansion-history comparison for selected CMB-$1000$ best fits.
The first three panels show every measured DESI DR2 transverse, radial, and
isotropic BAO/AP distance, respectively, normalized by the \LCDM\ best fit.
At the Ly$\alpha$ redshift the BAO-only point is
replaced by the recent joint BAO+AP compression
~\citep{DESI:2026lnd}.  The final panel shows the DES Dovekie
distance-modulus residuals~\citep{DES:2025sig}.  For visualization only, the
supernovae are binned following \citep{DESI:2025zgx}; the likelihood itself is evaluated using the
unbinned sample.}
\label{fig:expansion}
\end{figure*}

The result is different under the nonphantom prior.
The constraints become
$\wdm=0.001109^{+0.000514}_{-0.000502}$ and
$\wdm=0.001045^{+0.000448}_{-0.000444}$ for the two CMB treatments.
Zero is disfavored at approximately $2\sigma$ in both cases.
Equation~(\ref{eq:nonphantom}) removes the part of \CPL\ space that most
efficiently changes the intermediate-redshift distances through a phantom
phase.  The positive $\wdm$ dark-matter direction then again supplies
the needed separation between early and present matter densities.
Pad\'e-$w$+$\wdm$ gives a less significant but stable positive shift:
$\wdm=0.000644^{+0.000423}_{-0.000418}$ and
$\wdm=0.000544^{+0.000333}_{-0.000330}$, corresponding to about
$1.5\sigma$ and $1.6\sigma$, respectively.
Figure~\ref{fig:expansion} illustrates that the successful extensions make
coherent, modest changes to the distance--redshift relation.

\begin{table*}[htb]
\centering
\begin{tabular}{llc}
\toprule
CMB & model & $\wdm$\\
\midrule
CMB-$1000$ & \CPL+$\wdm$ &
$0.000272^{+0.000608}_{-0.000601}\;(-0.000013)$\\
& nonphantom \CPL+$\wdm$ &
$0.001109^{+0.000514}_{-0.000502}\;(0.001054)$\\
& Pad\'e-$w$+$\wdm$ &
$0.000644^{+0.000423}_{-0.000418}\;(0.000559)$\\
\midrule
CMB lensing-marginalized & \CPL+$\wdm$ &
$0.000276^{+0.000557}_{-0.000572}\;(0.000620)$\\
& nonphantom \CPL+$\wdm$ &
$0.001045^{+0.000448}_{-0.000444}\;(0.000740)$\\
& Pad\'e-$w$+$\wdm$ &
$0.000544^{+0.000333}_{-0.000330}\;(0.000528)$\\
\bottomrule
\end{tabular}
\caption{Dark-matter equation-of-state constraints in the dynamical models.
Each entry is the posterior mean with 68\% equal-tail errors, followed in
parentheses by the best fit.}
\label{tab:dynamicwdm}
\end{table*}

\section{Model comparison}
\label{sec:comparison}

Table~\ref{tab:modelcomparison} gives the model comparison.
We compare models through both $\Delta \chi^2$ and
\begin{equation}
 \DICV=\overline{\chi^2}+
 \frac{1}{2}{\rm Var}(\chi^2),
\label{eq:dicv}
\end{equation}
a variance-based deviance information criterion
~\citep{Spiegelhalter:2002yvw,Gelman2004BDA}.
All reported differences are relative to \LCDM\ unless explicitly stated.
The constant $ww$DM extension improves the best fit by
$\Delta\chi^2=-5.69$ and $-7.58$, while its $\DICV$ improves by
$-2.64$ and $-2.14$.
Neither one-parameter constant extension reproduces this pattern.
Unrestricted \CPL\ provides a larger improvement, confirming that the
distance data favor dynamical freedom even when dark matter is fixed.

The best overall values in this set are obtained by
Pad\'e-$w$+$\wdm$.
Relative to \LCDM, it gives
\begin{align}
\Delta\chi^2_{\rm Pad\acute e+DM-\Lambda CDM}&=(-9.25,-9.02),\nonumber\\
\Delta{\rm DIC}_{V,{\rm Pad\acute e+DM-\Lambda CDM}}&=(-6.29,-4.71),
\end{align}
where the entries refer to CMB-$1000$ and the CMB lensing-marginalized
likelihood.
Even relative to \CPL\ allowing phantom crossing, it retains an advantage in
both CMB treatments:
\begin{align}
\Delta\chi^2_{\rm Pad\acute e+DM-CPL}&=(-1.24,-1.10),\nonumber\\
\Delta{\rm DIC}_{V,{\rm Pad\acute e+DM-CPL}}&=(-0.89,-1.33).
\end{align}
The advantage over unrestricted \CPL+$\wdm$ and nonphantom
\CPL+$\wdm$ is larger in $\DICV$.
Accordingly, the comparison supports the joint nonphantom Pad\'e
description among the tested parameterizations.

\begin{table*}[htb]
\centering
\begin{tabular}{lrrrr}
\toprule
& \multicolumn{2}{c}{CMB-$1000$} &
\multicolumn{2}{c}{CMB lensing-marginalized}\\
model & $\Delta\chi^2$ & $\Delta\DICV$ &
$\Delta\chi^2$ & $\Delta\DICV$\\
\midrule
\LCDM & $0.00$ & $0.00$ & $0.00$ & $0.00$\\
$\Lambda w$DM & $-0.57$ & $+0.68$ & $-1.34$ & $+1.85$\\
$w$CDM & $-2.51$ & $-0.96$ & $-1.71$ & $+0.47$\\
$ww$DM & $-5.69$ & $-2.64$ & $-7.58$ & $-2.14$\\
\CPL & $-8.00$ & $-5.40$ & $-7.92$ & $-3.38$\\
\CPL+$\wdm$ & $-8.34$ & $-3.97$ & $-8.61$ & $-1.64$\\
nonphantom \CPL+$\wdm$ & $-6.60$ & $-2.13$ & $-6.73$ & $-1.40$\\
Pad\'e-$w$ & $-6.95$ & $-5.61$ & $-7.04$ & $-4.31$\\
Pad\'e-$w$+$\wdm$ & $\bm{-9.25}$ & $\bm{-6.29}$ &
$\bm{-9.02}$ & $\bm{-4.71}$\\
\bottomrule
\end{tabular}
\caption{Best-fit and variance-DIC differences relative to \LCDM.
Negative values favor the extension.}
\label{tab:modelcomparison}
\end{table*}

\section{Conclusions}
\label{sec:conclusions}

Although the standard cosmological model assumes the dark matter equation of state to be exactly zero, this remains an assumption rather than an established property of the dark sector. Since the microscopic nature of dark matter is still unknown, there is no fundamental reason why its effective equation of state must vanish identically. The dark matter sector may consist of multiple components with different physical properties, possess a small residual velocity dispersion, or include non-standard interactions or field dynamics, potentially leading to departures from the idealized pressureless limit (see also e.g. \cite{Luongo:2018lgy,Fichet:2023dju,Bottaro:2024pcb}). While current observations strongly favor a pressureless dark matter component, they do not require its effective equation of state to be exactly zero. It is therefore natural to test whether a small departure from $\wdm=0$ is compatible with current observations and can be further constrained by future surveys.

Furthermore, allowing the effective equation of state of dark matter to vary is also important because assumptions about the dark sector are inherently coupled when interpreting cosmological observations. A small departure from $\wdm=0$ modifies both the expansion history and the growth of cosmic structure, and may therefore be partially degenerate with the properties of dark energy. As a result, fixing the dark matter equation of state to be exactly zero could bias or artificially tighten inferences about whether dark energy evolves with time and about the significance of any deviation from the cosmological constant. This coupling is particularly relevant given the growing observational interest in evolving dark energy.

Motivated by these considerations, we have investigated models with a nonzero $\wdm$ using updated DESI DR2
distances, DES 5-year supernovae, and two complementary CMB likelihood
constructions.
Three conclusions follow.

First, when constant $\wdm$ and $\wde$ are released simultaneously, both
standard values are disfavored at approximately $2\sigma$ for either CMB
treatment.  Releasing either parameter alone produces no comparable effect.
Late-time BAO/AP and supernova geometry favor $\wde>-1$.
The CMB independently confines the matter-to-radiation ratio before recombination, equivalently the equality epoch $a_{eq}$ or early-Universe
$\Omega_mh^2$.  A positive $\wdm$ then changes the conversion between the
present and CMB-calibrated early dark-matter densities, allowing both sets
of constraints to be satisfied.

Second, the interpretation depends on phantom crossing.
Unrestricted \CPL+$\wdm$ makes dark matter consistent with the pressureless
limit because the crossing dark-energy history absorbs the geometric
freedom.  When crossing is forbidden, a positive $\wdm$ preference
persists.  It is approximately $2\sigma$ for nonphantom \CPL+$\wdm$ and
$1.5$--$1.6\sigma$ for Pad\'e-$w$, a flexible dark-energy parameterization
of quintessence evolution.

Third, for the baseline DES supernova sample, model comparisons based on
the best-fit $\chi^2$ and DIC mildly favor Pad\'e-$w$+$\wdm$ over
unrestricted \CPL\ in both CMB analyses, despite the latter allowing
phantom crossing whereas the former does not.  The apparent preference for
phantom crossing may therefore be a projection of departures from the
standard dark-matter sector.

The departures from \LCDM\ suggested by DESI should therefore be interpreted with caution, as they may reflect correlated physics across the dark sector rather than dark-energy evolution alone. Distinguishing these possibilities will require improved high-redshift observations, especially of Lyman-$\alpha$ tracers, together with more complete theoretical and observational tests of perturbations and structure formation.
Dedicated simulations and calibrated emulators, combined with observational tests from redshift-space distortions, galaxy clustering, weak lensing, and CMB lensing, will be crucial for determining whether the emerging geometric preference reflects new dark-energy dynamics, nonstandard dark-matter physics, or a correlated modification of both sectors.

\begin{acknowledgments}
We thank Eric Linder for helpful discussions. J.J. and A.S. acknowledge funding from the Korea Astronomy and Space Science Institute (KASI) through the project “Research on the Principles of the Accelerating Expansion of the Universe” (Project Code: 2026183201). A.S. is grateful for support by the Open KIAS Center at Korea Institute for Advanced Study. 
LLMs were used to improve the code and polish the language. The LLMs used included ChatGPT 5.3 to 5.6, DeepSeek V3 to V4 Flash, and GLM 3.5 Flash.
\end{acknowledgments}

\appendix

\onecolumngrid

\section{Priors and full posterior constraints}
\label{app:priors-posteriors}

Table~\ref{tab:appendix-priors} lists the cosmological and dark-sector
priors, with ${\cal U}(a,b)$ denoting a uniform distribution.
The neutrino sector is fixed with $\sum m_\nu=0.06\,{\rm eV}$.

\begin{table}[H]
\centering
\caption{Cosmological and dark-sector priors.}
\label{tab:appendix-priors}
\begin{tabular}{ll}
\toprule
Parameter & Prior\\
\midrule
$\ln(10^{10}A_s)$ & ${\cal U}(1.61,3.91)$\\
$n_s$ & ${\cal U}(0.8,1.2)$\\
$H_0\,[{\rm km\,s^{-1}\,Mpc^{-1}}]$ & ${\cal U}(20,100)$\\
$\Omega_b$ & ${\cal U}(0.03,0.07)$\\
$\Omega_{\rm DM}$ & ${\cal U}(0.001,0.99)$\\
$\tau_{\rm reio}$ & ${\cal U}(0.01,0.8)$\\
$\wdm$ & ${\cal U}(-0.5,0.3333)$\\
$\wde$ & ${\cal U}(-3,1)$\\
$w_{0,\rm DE}$ & ${\cal U}(-3,1)$\\
$w_{a,\rm DE}$ & ${\cal U}(-3,2)$\\
nonphantom CPL & $w_{0,\rm DE}>-1,\ 
w_{0,\rm DE}+w_{a,\rm DE}>-1$\\
$\epsilon_0$ & ${\cal U}(0,3)$\\
$\eta_0$ & ${\cal U}(0,100)$\\
\bottomrule
\end{tabular}
\end{table}

All posterior columns combine DESI DR2 and DES Dovekie supernovae with the
indicated CMB treatment.  Tables~\ref{tab:appendix-cpl-posteriors},
\ref{tab:appendix-pade-posteriors}, and
\ref{tab:appendix-constant-posteriors} give posterior means and 68\%
equal-tail errors; the values in parentheses are the best fits.

\begin{table}[H]
\centering
\footnotesize
\setlength{\tabcolsep}{4pt}
\caption{Full posterior constraints for the CPL-based models.
Each entry is the posterior mean with 68\%
equal-tail errors, followed in parentheses by the best fit.}
\label{tab:appendix-cpl-posteriors}
\begin{tabular}{lccc}
\toprule
Parameter & \CPL & \CPL+$\wdm$ & nonphantom-\CPL+$\wdm$\\
\midrule
\multicolumn{4}{l}{\textit{CMB-$1000$}}\\
$\ln(10^{10}A_s)$ & $3.028^{+0.017}_{-0.017}\;(3.033)$ & $3.029^{+0.017}_{-0.017}\;(3.027)$ & $3.031^{+0.017}_{-0.017}\;(3.023)$\\
$n_s$ & $0.9678^{+0.0056}_{-0.0057}\;(0.9689)$ & $0.9677^{+0.0056}_{-0.0057}\;(0.9682)$ & $0.9698^{+0.0057}_{-0.0058}\;(0.9669)$\\
$H_0$ & $67.20^{+0.55}_{-0.55}\;(67.28)$ & $67.23^{+0.55}_{-0.54}\;(67.07)$ & $67.41^{+0.53}_{-0.53}\;(67.34)$\\
$\Omega_b$ & $0.04934^{+0.00083}_{-0.00084}\;(0.04926)$ & $0.04923^{+0.00091}_{-0.00090}\;(0.04959)$ & $0.04891^{+0.00089}_{-0.00088}\;(0.04889)$\\
$\Omega_{\rm DM}$ & $0.2618^{+0.0049}_{-0.0049}\;(0.2613)$ & $0.2605^{+0.0058}_{-0.0058}\;(0.2622)$ & $0.2536^{+0.0045}_{-0.0045}\;(0.2559)$\\
$\tau_{\rm reio}$ & $0.0496^{+0.0077}_{-0.0077}\;(0.0476)$ & $0.0498^{+0.0078}_{-0.0077}\;(0.0488)$ & $0.0497^{+0.0079}_{-0.0077}\;(0.0484)$\\
$\wdm$ & $0\ \mathrm{(fixed)}$ & $0.000272^{+0.000608}_{-0.000601}\;(-0.000013)$ & $0.001109^{+0.000514}_{-0.000502}\;(0.001054)$\\
$w_{0,\rm DE}$ & $-0.8395^{+0.0562}_{-0.0562}\;(-0.8387)$ & $-0.8506^{+0.0609}_{-0.0613}\;(-0.8399)$ & $-0.9356^{+0.0334}_{-0.0342}\;(-0.9109)$\\
$w_{a,\rm DE}$ & $-0.543^{+0.231}_{-0.230}\;(-0.550)$ & $-0.456^{+0.296}_{-0.290}\;(-0.503)$ & $0.030^{+0.096}_{-0.090}\;(-0.089)$\\
\midrule
\multicolumn{4}{l}{\textit{CMB lensing-marginalized}}\\
$\ln(10^{10}A_s)$ & $3.029^{+0.017}_{-0.017}\;(3.022)$ & $3.030^{+0.017}_{-0.017}\;(3.027)$ & $3.031^{+0.017}_{-0.016}\;(3.020)$\\
$n_s$ & $0.9726^{+0.0039}_{-0.0039}\;(0.9738)$ & $0.9730^{+0.0039}_{-0.0039}\;(0.9735)$ & $0.9752^{+0.0039}_{-0.0038}\;(0.9746)$\\
$H_0$ & $67.29^{+0.54}_{-0.54}\;(67.29)$ & $67.35^{+0.56}_{-0.56}\;(67.42)$ & $67.49^{+0.55}_{-0.54}\;(67.06)$\\
$\Omega_b$ & $0.04953^{+0.00081}_{-0.00081}\;(0.04945)$ & $0.04942^{+0.00085}_{-0.00085}\;(0.04927)$ & $0.04919^{+0.00082}_{-0.00083}\;(0.04985)$\\
$\Omega_{\rm DM}$ & $0.2614^{+0.0045}_{-0.0046}\;(0.2614)$ & $0.2595^{+0.0059}_{-0.0058}\;(0.2576)$ & $0.2530^{+0.0047}_{-0.0048}\;(0.2574)$\\
$\tau_{\rm reio}$ & $0.0494^{+0.0081}_{-0.0080}\;(0.0491)$ & $0.0498^{+0.0081}_{-0.0080}\;(0.0473)$ & $0.0501^{+0.0080}_{-0.0079}\;(0.0476)$\\
$\wdm$ & $0\ \mathrm{(fixed)}$ & $0.000276^{+0.000557}_{-0.000572}\;(0.000620)$ & $0.001045^{+0.000448}_{-0.000444}\;(0.000740)$\\
$w_{0,\rm DE}$ & $-0.8434^{+0.0547}_{-0.0538}\;(-0.8722)$ & $-0.8590^{+0.0625}_{-0.0625}\;(-0.8723)$ & $-0.9352^{+0.0342}_{-0.0354}\;(-0.8956)$\\
$w_{a,\rm DE}$ & $-0.511^{+0.213}_{-0.211}\;(-0.401)$ & $-0.405^{+0.303}_{-0.303}\;(-0.289)$ & $0.041^{+0.107}_{-0.100}\;(-0.093)$\\
\bottomrule
\end{tabular}
\end{table}

\begin{table}[H]
\centering
\footnotesize
\setlength{\tabcolsep}{6pt}
\caption{Full posterior constraints for the Pad\'e-based models.  Each
entry is the posterior mean with 68\% equal-tail errors, followed in
parentheses by the best fit.}
\label{tab:appendix-pade-posteriors}
\begin{tabular}{lcc}
\toprule
Parameter & Pad\'e-$w$ & Pad\'e-$w$+$\wdm$\\
\midrule
\multicolumn{3}{l}{\textit{CMB-$1000$}}\\
$\ln(10^{10}A_s)$ & $3.029^{+0.017}_{-0.017}\;(3.032)$ & $3.030^{+0.017}_{-0.016}\;(3.038)$\\
$n_s$ & $0.9722^{+0.0050}_{-0.0049}\;(0.9732)$ & $0.9685^{+0.0056}_{-0.0056}\;(0.9695)$\\
$H_0$ & $66.59^{+0.69}_{-0.68}\;(66.55)$ & $66.77^{+0.68}_{-0.67}\;(66.99)$\\
$\Omega_b$ & $0.05055^{+0.00103}_{-0.00106}\;(0.05064)$ & $0.04988^{+0.00110}_{-0.00110}\;(0.04970)$\\
$\Omega_{\rm DM}$ & $0.2633^{+0.0057}_{-0.0057}\;(0.2636)$ & $0.2616^{+0.0056}_{-0.0056}\;(0.2595)$\\
$\tau_{\rm reio}$ & $0.0513^{+0.0078}_{-0.0077}\;(0.0535)$ & $0.0497^{+0.0078}_{-0.0077}\;(0.0503)$\\
$\wdm$ & $0\ \mathrm{(fixed)}$ & $0.000644^{+0.000423}_{-0.000418}\;(0.000559)$\\
$\epsilon_0$ & $1.011^{+0.522}_{-0.530}\;(1.301)$ & $1.134^{+0.550}_{-0.554}\;(0.975)$\\
$\eta_0$ & $64.50^{+26.52}_{-28.50}\;(90.73)$ & $64.49^{+26.35}_{-27.35}\;(60.19)$\\
\midrule
\multicolumn{3}{l}{\textit{CMB lensing-marginalized}}\\
$\ln(10^{10}A_s)$ & $3.027^{+0.017}_{-0.016}\;(3.017)$ & $3.031^{+0.017}_{-0.016}\;(3.011)$\\
$n_s$ & $0.9754^{+0.0037}_{-0.0036}\;(0.9760)$ & $0.9738^{+0.0038}_{-0.0038}\;(0.9736)$\\
$H_0$ & $66.65^{+0.64}_{-0.65}\;(66.39)$ & $66.86^{+0.69}_{-0.68}\;(67.13)$\\
$\Omega_b$ & $0.05060^{+0.00101}_{-0.00099}\;(0.05091)$ & $0.05015^{+0.00105}_{-0.00105}\;(0.04971)$\\
$\Omega_{\rm DM}$ & $0.2636^{+0.0054}_{-0.0053}\;(0.2648)$ & $0.2614^{+0.0056}_{-0.0056}\;(0.2592)$\\
$\tau_{\rm reio}$ & $0.0504^{+0.0081}_{-0.0079}\;(0.0488)$ & $0.0499^{+0.0081}_{-0.0079}\;(0.0438)$\\
$\wdm$ & $0\ \mathrm{(fixed)}$ & $0.000544^{+0.000333}_{-0.000330}\;(0.000528)$\\
$\epsilon_0$ & $1.001^{+0.516}_{-0.515}\;(1.197)$ & $1.155^{+0.555}_{-0.560}\;(1.209)$\\
$\eta_0$ & $65.34^{+26.04}_{-28.17}\;(66.12)$ & $64.91^{+26.24}_{-28.15}\;(90.63)$\\
\bottomrule
\end{tabular}
\end{table}

\begin{table}[H]
\centering
\footnotesize
\setlength{\tabcolsep}{3.5pt}
\caption{Full posterior constraints for the constant dark-sector models.
Each entry is the posterior mean with 68\% equal-tail errors, followed in
parentheses by the best fit.}
\label{tab:appendix-constant-posteriors}
\begin{tabular}{lcccc}
\toprule
Parameter & \LCDM & $\Lambda w$DM & $w$CDM & $ww$DM\\
\midrule
\multicolumn{5}{l}{\textit{CMB-$1000$}}\\
$\ln(10^{10}A_s)$ & $3.028^{+0.018}_{-0.017}\;(3.032)$ & $3.029^{+0.017}_{-0.017}\;(3.027)$ & $3.028^{+0.017}_{-0.017}\;(3.038)$ & $3.031^{+0.017}_{-0.017}\;(3.035)$\\
$n_s$ & $0.9707^{+0.0049}_{-0.0050}\;(0.9714)$ & $0.9682^{+0.0057}_{-0.0057}\;(0.9680)$ & $0.9734^{+0.0053}_{-0.0053}\;(0.9731)$ & $0.9690^{+0.0058}_{-0.0058}\;(0.9700)$\\
$H_0$ & $68.18^{+0.29}_{-0.29}\;(68.27)$ & $68.41^{+0.39}_{-0.39}\;(68.45)$ & $67.47^{+0.51}_{-0.53}\;(67.45)$ & $67.51^{+0.54}_{-0.55}\;(67.73)$\\
$\Omega_b$ & $0.04809^{+0.00032}_{-0.00032}\;(0.04807)$ & $0.04755^{+0.00070}_{-0.00070}\;(0.04753)$ & $0.04927^{+0.00079}_{-0.00078}\;(0.04930)$ & $0.04877^{+0.00091}_{-0.00090}\;(0.04838)$\\
$\Omega_{\rm DM}$ & $0.2524^{+0.0033}_{-0.0034}\;(0.2515)$ & $0.2507^{+0.0039}_{-0.0039}\;(0.2501)$ & $0.2559^{+0.0040}_{-0.0041}\;(0.2560)$ & $0.2539^{+0.0042}_{-0.0042}\;(0.2524)$\\
$\tau_{\rm reio}$ & $0.0505^{+0.0082}_{-0.0077}\;(0.0502)$ & $0.0496^{+0.0078}_{-0.0076}\;(0.0487)$ & $0.0512^{+0.0079}_{-0.0079}\;(0.0484)$ & $0.0497^{+0.0078}_{-0.0077}\;(0.0475)$\\
$\wdm$ & $0\ \mathrm{(fixed)}$ & $0.000370^{+0.000417}_{-0.000424}\;(0.000358)$ & $0\ \mathrm{(fixed)}$ & $0.000968^{+0.000501}_{-0.000496}\;(0.000913)$\\
$\wde$ & $-1\ \mathrm{(fixed)}$ & $-1\ \mathrm{(fixed)}$ & $-0.9649^{+0.0215}_{-0.0214}\;(-0.9645)$ & $-0.9380^{+0.0259}_{-0.0262}\;(-0.9481)$\\
\midrule
\multicolumn{5}{l}{\textit{CMB lensing-marginalized}}\\
$\ln(10^{10}A_s)$ & $3.028^{+0.016}_{-0.016}\;(3.019)$ & $3.029^{+0.016}_{-0.016}\;(3.025)$ & $3.027^{+0.017}_{-0.016}\;(3.027)$ & $3.031^{+0.017}_{-0.017}\;(3.026)$\\
$n_s$ & $0.9743^{+0.0035}_{-0.0035}\;(0.9735)$ & $0.9733^{+0.0037}_{-0.0037}\;(0.9720)$ & $0.9757^{+0.0038}_{-0.0037}\;(0.9723)$ & $0.9746^{+0.0037}_{-0.0038}\;(0.9756)$\\
$H_0$ & $68.27^{+0.25}_{-0.25}\;(68.37)$ & $68.53^{+0.38}_{-0.39}\;(68.70)$ & $67.58^{+0.54}_{-0.55}\;(67.60)$ & $67.57^{+0.51}_{-0.52}\;(67.65)$\\
$\Omega_b$ & $0.04817^{+0.00031}_{-0.00031}\;(0.04815)$ & $0.04774^{+0.00059}_{-0.00058}\;(0.04742)$ & $0.04924^{+0.00081}_{-0.00081}\;(0.04922)$ & $0.04909^{+0.00079}_{-0.00078}\;(0.04906)$\\
$\Omega_{\rm DM}$ & $0.2523^{+0.0030}_{-0.0031}\;(0.2513)$ & $0.2503^{+0.0039}_{-0.0038}\;(0.2485)$ & $0.2560^{+0.0041}_{-0.0041}\;(0.2563)$ & $0.2536^{+0.0041}_{-0.0040}\;(0.2535)$\\
$\tau_{\rm reio}$ & $0.0501^{+0.0079}_{-0.0077}\;(0.0478)$ & $0.0495^{+0.0079}_{-0.0078}\;(0.0499)$ & $0.0504^{+0.0081}_{-0.0078}\;(0.0517)$ & $0.0498^{+0.0081}_{-0.0080}\;(0.0516)$\\
$\wdm$ & $0\ \mathrm{(fixed)}$ & $0.000294^{+0.000324}_{-0.000323}\;(0.000425)$ & $0\ \mathrm{(fixed)}$ & $0.000870^{+0.000408}_{-0.000410}\;(0.000707)$\\
$\wde$ & $-1\ \mathrm{(fixed)}$ & $-1\ \mathrm{(fixed)}$ & $-0.9687^{+0.0216}_{-0.0216}\;(-0.9719)$ & $-0.9353^{+0.0258}_{-0.0254}\;(-0.9435)$\\
\bottomrule
\end{tabular}
\end{table}

\clearpage
\twocolumngrid

\section{Alternative supernova samples}
\label{app:alternative-sn}

We test alternative Type Ia supernova compilation by replacing
the baseline DES Dovekie likelihood~\citep{DES:2025sig} with
Pantheon+~\citep{Brout:2022vxf} or Union3~\citep{Rubin:2023jdq}.  Here we
restrict the test to CMB-$1000$, always combined with DESI DR2.

\begin{figure*}[htb]
\centering
\includegraphics[width=0.82\textwidth]{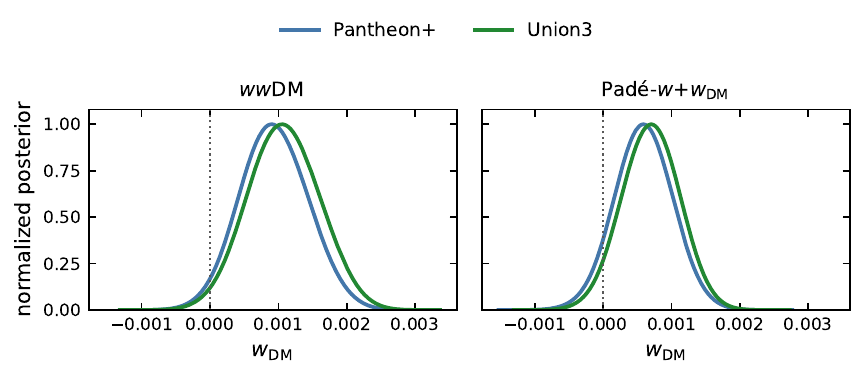}
\caption{Alternative-supernova constraints using DESI DR2 and CMB-$1000$.
The panels show the marginalized $\wdm$ posteriors for $ww$DM (left) and
Pad\'e-$w$+$\wdm$ (right), comparing Pantheon+ with Union3.  The dotted
line marks $\wdm=0$.}
\label{fig:alternative-sn}
\end{figure*}

Figure~\ref{fig:alternative-sn} compares the two models in this test with
free $\wdm$.  In $ww$DM, Pantheon+ gives
\begin{equation}
\begin{aligned}
\wdm&=0.000939^{+0.000503}_{-0.000498},
\end{aligned}
\end{equation}
Union3 shifts the mean slightly upward,
\begin{equation}
\begin{aligned}
\wdm&=0.001078^{+0.000533}_{-0.000520}.
\end{aligned}
\end{equation}
Thus the positive $ww$DM trend persists under
both replacements.

For Pad\'e-$w$+$\wdm$, Pantheon+ gives
\begin{equation}
\begin{aligned}
\wdm&=0.000592^{+0.000419}_{-0.000421},
\end{aligned}
\end{equation}
while Union3 gives
\begin{equation}
\begin{aligned}
\wdm&=0.000696^{+0.000422}_{-0.000424}.
\end{aligned}
\end{equation}
Both posterior
means remain positive, although zero is contained in their 95\% intervals.

Both of them improve over \LCDM.  For Pantheon+, relative to \LCDM,
$ww$DM gives $\Delta\chi^2=-4.46$ and $\Delta\DICV=-1.14$, while
Pad\'e-$w$+$\wdm$ gives $\Delta\chi^2=-5.39$ and
$\Delta\DICV=-2.95$.  For Union3, $ww$DM gives
$\Delta\chi^2=-5.64$ and $\Delta\DICV=-2.41$, while
Pad\'e-$w$+$\wdm$ gives $\Delta\chi^2=-9.88$ and
$\Delta\DICV=-7.87$.
Comparing Pad\'e-$w$+$\wdm$ and unrestricted \CPL\ gives
\begin{align}
\Delta\chi^2_{\rm Pad\acute e+DM-CPL}
&=(-0.014,+0.539),\nonumber\\
\Delta{\rm DIC}_{V,{\rm Pad\acute e+DM-CPL}}
&=(-0.301,-0.492),
\end{align}
where the entries refer to Pantheon+ and Union3.  Pad\'e-$w$+$\wdm$
therefore retains a $\DICV$ advantage over unrestricted \CPL\ for both
alternative supernova samples.  Pantheon+ also gives a slightly better
best fit, while Union3 reverses the best-fit ordering slightly.

\section{Nonlinear calculation}
\label{app:nonlinear}

Our analysis avoids relying on a calibrated
nonlinear matter-power prescription for $\wdm\ne0$.
We instead use the following calculation only to diagnose the size and
direction of nonlinear effects for the $ww$DM model.  For every sampled cosmology, the modified
\texttt{CLASS} gives the transfer function for the
clustering-matter contrast
\begin{equation}
 \delta_{\rm cl}(k,a)=
 \frac{\rho_b(a)\delta_b(k,a)+\rho_{\rm DM}(a)\delta_{\rm DM}(k,a)}
 {\rho_b(a)+\rho_{\rm DM}(a)} .
\label{eq:deltacl}
\end{equation}
The corresponding linear spectrum is constructed directly from the
primordial curvature spectrum,
\begin{equation}
 P_{\rm cl}^{\rm lin}(k,a)=\frac{2\pi^2}{k^3}
 A_s\left(\frac{k}{k_*}\right)^{n_s-1}
 T_{\rm cl}^2(k,a).
\label{eq:pcllin}
\end{equation}
The linear growth factor $D(a)$ is measured from
this spectrum at $k=0.1\,h\,{\rm Mpc}^{-1}$ and normalized to unity today.
This reference scale lies well inside the horizon throughout the collapse
integration.

For spherical collapse we retain independent baryon and DM velocities,
as required for a multicomponent top hat~\citep{Pace:2019vrs}.  Define
\begin{equation}
 s_i=\ln(1+\delta_i^{\rm th}),\qquad
 q_i=\frac{\theta_i}{aH},\qquad
 {\cal O}_i(a)=\frac{\rho_i(a)}{\rho_{\rm crit,0}},
\end{equation}
where $\theta_i$ is the conformal velocity divergence and primes below
denote derivatives with respect to $a$.  For pressureless baryons and
$c_{s,\rm DM}^2=0$, the nonlinear continuity and Euler equations become
\begin{align}
 s_b'={}&-\frac{q_b}{a},\nonumber\\
 s_{\rm DM}'={}&\frac{3\wdm}{a}(1-e^{-s_{\rm DM}})
 -\frac{q_{\rm DM}}{a}(1+\wdm e^{-s_{\rm DM}}),\label{eq:topdeltas}\\
 q_i'={}&-\left(\frac{2}{a}+\frac{d\ln H}{da}\right)q_i
 -\frac{q_i^2}{3a}\nonumber\\
 &-\frac{3}{2aE^2(a)}
 \sum_{j=b,{\rm DM}}{\cal O}_j(a)(e^{s_j}-1),
\label{eq:collapse}
\end{align}
with $E(a)=H(a)/H_0$.

At $a_i=0.005$ we initialize the linear density and velocity ratios directly
from the \texttt{CLASS} density and velocity-divergence transfers.  If
$A_i$ is the initial clustering-matter amplitude and all transfers are
evaluated at $k_{\rm ref}=0.1\,h\,{\rm Mpc}^{-1}$, then
\begin{align}
 s_{j,i}&=\ln\left[1+A_i\frac{T_j(a_i)}{T_{\rm cl}(a_i)}\right],\nonumber\\
 q_{j,i}&=\frac{A_i T_{\theta_j}(a_i)}
 {a_iH(a_i)T_{\rm cl}(a_i)},\qquad j\in\{b,{\rm DM}\}.
\end{align}

We tune $A_i$ until $s_{\rm DM}=-3\ln(10^{-4})$ at the target collapse
scale factor $a_c$, using this large-density value as the asymptotic-collapse
convention, and define the reference-scale linearly extrapolated threshold
\begin{equation}
 \delta_c(a_c)=A_i
 \frac{T_{\rm cl}(k_{\rm ref},a_c)}
 {T_{\rm cl}(k_{\rm ref},a_i)} .
\label{eq:deltac}
\end{equation}
DM-flow turnaround is located by $q_{\rm DM}=-3$.  If $a_{\rm ta}$ denotes
that solution, the virial overdensity used in the halo calculation is the
density ratio obtained with the approximate physical-radius closure
$R_{\rm vir}^{\rm phys}=R_{\rm ta}^{\rm phys}/2$,
\begin{equation}
 \begin{split}
 \Delta_{\rm vir}=\frac{8}{{\cal O}_{\rm cl}(a_c)}
 \bigl\{&
 {\cal O}_b(a_{\rm ta})[1+\delta_b^{\rm th}(a_{\rm ta})]\\
 &+{\cal O}_{\rm DM}(a_{\rm ta})
 [1+\delta_{\rm DM}^{\rm th}(a_{\rm ta})]\bigr\}.
 \end{split}
\label{eq:deltavir}
\end{equation}
The factor of eight is a deliberately dust-like virial closure.

The mass variance is evaluated with a real-space top-hat window,
\begin{align}
 \sigma^2(M,a)&=\frac{1}{2\pi^2}\int dk\,k^2P_{\rm cl}^{\rm lin}(k,a)
 W^2(kR_L),\\
 W(y)&=3\frac{\sin y-y\cos y}{y^3},\qquad
 R_L=\left(\frac{3M}{4\pi\rho_{\rm cl,0}}\right)^{1/3}.
\end{align}
Here $M$ is a present-day clustering-energy-equivalent Lagrangian mass label.
Because $a^3\rho_{\rm cl}(a)$ is not constant for $\wdm\ne0$, it is not
identified with an exactly conserved particle rest mass.
With $\nu=\delta_c(a)/\sigma(M,a)$, we use the Press--Schechter mass
function and its associated linear bias~\citep{Press:1973iz},
\begin{align}
 \frac{dn}{dM}&=\sqrt{\frac{2}{\pi}}\frac{\rho_{\rm cl,0}}{M^2}
 \nu\left|\frac{d\ln\sigma}{d\ln M}\right|e^{-\nu^2/2},\\
 b(M)&=1+\frac{\nu^2-1}{\delta_c}.
\end{align}
The mass function and bias are renormalized numerically so that
$\int dM\,M(dn/dM)/\rho_{\rm cl,0}=1$ and its bias-weighted counterpart is
also unity.  Each halo is represented by a top-hat profile, while the
one-halo term uses its Lagrangian compensation,
\begin{align}
 u(k|M)&=W(kr_{\rm vir}^{\rm prof}),\\
 u_{\rm c}(k|M)&=W(kr_{\rm vir}^{\rm prof})-W(kR_L),
\end{align}
\begin{equation}
 r_{\rm vir}^{\rm prof}=\left(\frac{3M}
 {4\pi\Delta_{\rm vir}\rho_{\rm cl,0}}\right)^{1/3}.
\end{equation}
The latter is a comoving profile-radius convention associated with our
present-day energy-equivalent mass label; for $\wdm\ne0$ it is not
identified with the exact ratio $R_{\rm vir}^{\rm phys}/a_c$.
The one- and two-halo terms are then~\citep{Cooray:2002dia}
\begin{align}
 P_{\rm 1h}(k)&=\int dM\,\frac{dn}{dM}
 \left(\frac{M}{\rho_{\rm cl,0}}\right)^2u_{\rm c}^2(k|M),\\
 P_{\rm 2h}(k)&=P_{\rm cl}^{\rm lin}(k)
 \left[\int dM\,\frac{dn}{dM}b(M)
 \frac{M}{\rho_{\rm cl,0}}u(k|M)\right]^2 .
\end{align}

\begin{figure*}[htb]
\centering
\includegraphics[width=0.58\textwidth]{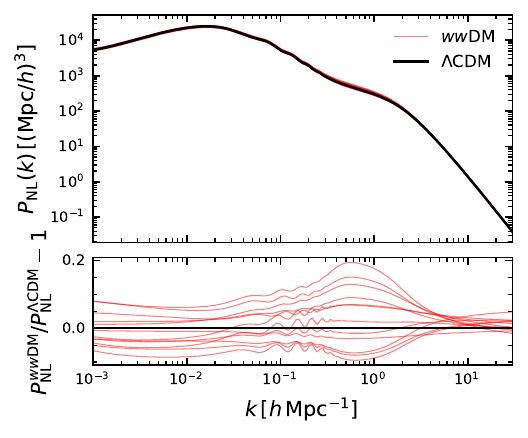}
\caption{Nonlinear diagnostic at $z=0$ for draws from the current
CMB-$1000$ $ww$DM chain.  Thin curves show individual posterior draws and
the lower panel gives ratios to the \LCDM\ best fit.  This spherical-collapse
halo model is not used in the likelihood and should be interpreted as an
order-of-magnitude sensitivity test, not a precision nonlinear prediction.}
\label{fig:nonlinear}
\end{figure*}

Fig.~\ref{fig:nonlinear} shows the $z=0$ nonlinear matter power spectrum results for 12 weighted draws from the
CMB-$1000$ $ww$DM posterior, compared with the
\LCDM\ best fit.
We find no coherent, statistically significant deviation of the $ww$DM
ensemble from \LCDM.  This diagnostic, however, cannot replace a dedicated
simulation-calibrated emulator, which will be required when incorporating
small-scale information.

\section{Applicability of the Lyman-$\alpha$ AP compression}
\label{app:lya-ap-validation}
The Lyman-$\alpha$ broadband AP measurement used in this work is obtained from a full-shape fit that adopts a fiducial \LCDM\ linear power spectrum~\citep{DESI:2026lnd}. It is therefore not, in a strict sense, independent of the assumed clustering template.

In our generalized dark matter implementation the rest-frame sound speed is fixed to $c_{s,\rm DM}^2=0$. Consequently, a small nonzero $\wdm$ does not generate a pressure-supported Jeans scale of the type that would arise for $c_{s,\rm DM}^2>0$. Nevertheless, it can modify the smooth broadband shape and growth history, so the absence of such a scale alone may not be sufficient to guarantee the validity of a \LCDM-based full-shape compression.
In the DESI Lyman-$\alpha$ analysis, much of this freedom is marginalized through the smooth isotropic dilation, tracer biases, redshift-space-distortion parameters, and other nuisance parameters.
We therefore test explicitly whether the residual mismatch between a $\wdm\ne0$  spectrum and the fiducial \LCDM\ template biases the recovered broadband AP parameter.

The relevant anisotropic parameter can be written as
\begin{equation}
\phi_s =
\frac{q_\perp}{q_\parallel}=\frac{(D_M/D_H)}{[D_M/D_H]_{\rm fid}},
\label{eq:phis-ap}
\end{equation}
where the subscript $s$ denotes the smooth broadband component. The corresponding smooth isotropic scaling $\alpha_s$ is marginalized and is not used as a cosmological observable.

Following \cite{DESI:2026lnd,Herbold:2026syy}, we perform a noiseless injection--recovery test using the same Lyman-$\alpha$ correlation-function modeling implemented in \texttt{Vega}~\citep{Cuceu:2022brl}. As the injected cosmology we take a representative best-fitting $ww$DM model from our analysis.
We generate the matter power spectrum at $z_{\rm eff}=2.33$.
This spectrum, together with the background distances of the same cosmology, is propagated through the Lyman-$\alpha$ forest auto-correlation and Lyman-$\alpha$--quasar cross-correlation models to construct a noiseless synthetic data vector.

The synthetic correlations are then analyzed with the DESI fiducial \LCDM\ power-spectrum template rather than with the spectrum used for their generation. 
For the likelihood weighting we use the publicly available smoothed DESI DR1 covariance~\citep{DESI:2024lzq}, retaining the Region A Lyman-$\alpha$ auto-correlation and quasar--Lyman-$\alpha$ cross-correlation blocks used in the test (following \cite{Herbold:2026syy}).
We apply a single overall rescaling of this covariance.
The rescaling is calibrated with a fiducial-\LCDM\ injection and recovery such that the resulting uncertainty is $\sigma_{\phi_s}\simeq0.011$, matching the DR2 measurement, and the same rescaled covariance is then held fixed for the $ww$DM test.

For this test the input geometry is
\begin{equation}
\phi_s^{\rm true}=0.989091 .
\end{equation}
The fit using the fiducial \LCDM\ template gives
\begin{equation}
\phi_s^{\rm rec}=0.989159,
\end{equation}
and hence
\begin{equation}
\Delta\phi_s\equiv
\phi_s^{\rm rec}-\phi_s^{\rm true}
=6.8\times10^{-5}.
\label{eq:phis-bias}
\end{equation}
Relative to the DESI DR2 statistical uncertainty,
$\sigma_{\phi_s}\simeq0.011$, this corresponds to only
\begin{equation}
\frac{|\Delta\phi_s|}{\sigma_{\phi_s}}
\simeq 0.006 ,
\end{equation}
well below the $1/3 \sigma$ criterion used in the DESI validation tests~\citep{Herbold:2026syy}.
Thus, the mismatch
with the fiducial \LCDM\ template induces a negligible shift in the
recovered broadband AP parameter at the current precision.
This result supports the use of the Lyman-$\alpha$ AP compression in our
analysis.

\bibliography{refs}{}
\bibliographystyle{aasjournalv7.1}

\end{CJK*}
\end{document}